\documentclass[onecolumn,journal]{IEEEtran}

\usepackage{amsmath,amsxtra,amssymb,amsthm,amsfonts,eufrak,bm,dsfont, flushend}
\usepackage[usenames,dvipsnames]{xcolor}
\usepackage{colortbl,booktabs,arydshln,multirow}
\usepackage{hyperref}
\usepackage{cleveref}
\usepackage{diagbox}
\usepackage[noadjust]{cite}
\usepackage{tikz-cd}
\usepackage{float}
\usepackage{mathrsfs}
\usetikzlibrary{matrix,arrows,decorations.pathmorphing}
\usepackage{soul}
\usepackage{bbm}

\newtheorem{lemma}{Lemma}

\newtheorem{theorem}[lemma]{Theorem}

\newtheorem{rem}[lemma]{Remark}

\newcommand{\re}{\begin{rem}\rm}
\newcommand{\mar}{\end{rem}}
\newtheorem{exam}[lemma]{Example}

\newcommand{\qd}{\end{proof}\vspace{0.5ex}}
\newcommand{\prf}{\begin{proof}[\bf Proof:]}

\newcommand{\pl}{\hspace{.1cm}}

\renewcommand{\H}{{\mathcal H}}

\newcommand{\N}{{\mathcal N}}

\newcommand{\eps}{\varepsilon}

\newcommand{\bra}[1]{\langle{#1}|}
\newcommand{\ket}[1]{|{#1}\rangle}
\newcommand{\ketbra}[1]{|{#1}\rangle\langle{#1}|}

\newcommand{\ten}{\otimes}

\newcommand{\tr}{{\text{tr}}}

\renewcommand{\H}{\mathcal{H}}

\newcommand{\cH}{\mathcal{H}}

\allowdisplaybreaks

\begin{document}

\title{Adversarial Hypothesis Testing \\for Quantum Channels}

\author{Masahito Hayashi, \IEEEmembership{Fellow, IEEE}, Hao-Chung Cheng, and Li Gao
\thanks{Masahito Hayashi is with
School of Data Science, The Chinese University of Hong Kong,
Shenzhen, Longgang District, Shenzhen, 518172, China,
International Quantum Academy, Futian District, Shenzhen 518048, China,
and
the Graduate School of Mathematics, Nagoya University, Nagoya, 464-8602, Japan
(E-mail: \texttt{\href{mailto:hmasahito@cuhk.edu.cn}{hmasahito@cuhk.edu.cn}}
).}
\thanks{
Hao-Chung Cheng is with the Department of Electrical Engineering, the Graduate Institute of Communication Engineering, the Department of Mathematics, and the Institute of Applied Mathematical Sciences, and the Center for Quantum Science and Engineering, National Taiwan University, Taipei 10617, Taiwan,
and the Physics Division, National Center for Theoretical Sciences, Taipei 10617, Taiwan
and with the Hon Hai (Foxconn) Quantum Computing Center, New Taipei City 236, Taiwan
(E-mail: \texttt{\href{mailto:haochung@ntu.edu.tw}{haochung@ntu.edu.tw}}).}
\thanks{
Li Gao is with
the  School of Mathematics and Statistics, Wuhan University, Wuhan, 430072, China
and
Wuhan Institute of Quantum Science and Technology, Wuhan 430075, China
(E-mail: \texttt{\href{mailto:gao.li@whu.edu.cn}{gao.li@whu.edu.cn}}
).}
}
\markboth{M.~Hayashi, H.-C.~Cheng, and
L.~Gao: Adversarial Hypothesis Testing}{}

\maketitle

\begin{abstract}
This paper presents a systematic study of adversarial hypothesis testing for both quantum-quantum (QQ) and classical-quantum (CQ) channels. 
Unlike conventional channel discrimination, we consider a framework where the sender, Alice, selects the channel input adversarially to minimize Bob's distinguishability. 
We analyze this problem across four settings based on whether Alice employs i.i.d.~or general inputs and whether the receiver, Bob, is informed of the specific input choice (allowing his measurement to depend on the input).
We characterize the Stein exponents for each setting and reveal a striking distinction in behavior: for QQ channels with i.i.d.~inputs, Bob’s knowledge of the input significantly enhances distinguishability, yet this advantage vanishes when general inputs are permitted. 
In contrast, for CQ channels, Bob being informed provides a consistent advantage over the corresponding entanglement-breaking channels for both i.i.d.~and general inputs. 
These results demonstrate a unique phenomenon in adversarial hypothesis testing where the CQ channel does not merely behave as a special case of the QQ channel.
\end{abstract}


\section{Introduction} \label{sec:introduction}

The binary hypothesis testing considers the case that given the system is in one of the two states, $\rho$ as the null hypothesis and $\sigma$ as the alternative hypothesis, one wants to distinguish $\rho$ from $\sigma$ via implementing some test operator $0\le T\le I$. 
The type I error $\alpha$ and type II error $\beta$ are defined as
$\alpha=\tr((I-T)\rho))$, $\beta=\tr(T\sigma)$.
In the independent and identically distributed (i.i.d.) setting for distinguishing $\rho^{\otimes n}$ from $\sigma^{\otimes n}$, Stein's Lemma provides a fundamental \emph{distinguishability} for the hypothesis testing \cite{HP91, ON00}.
That is, the largest asymptotic exponential decaying rate of $\beta$, while $\alpha$ is at most a constant $\varepsilon \in (0,1)$, is given by Umegaki's relative entropy $D(\rho\|\sigma)$ \cite{Ume62}.

We consider binary hypothesis testing for channels.
Let $\mathcal{N}_1,\mathcal{N}_2$ be two quantum channels from Alice's system to Bob's. The setting of the problem is: the sender Alice sends an input state $\rho$,
through the channel $\mathcal{N}_1$ (the  null hypothesis), or the channel $\mathcal{N}_2$
(the alternative hypothesis). Then the receiver Bob will implement some measurement $T$ on the output, which will be either $\mathcal{N}_1(\rho)$ or $\mathcal{N}_2(\rho)$, to distinguish the case $\mathcal{N}_1$ being used from that $\mathcal{N}_2$ is used.
In this work, Alice is \emph{adversarial} in the sense that she will choose the \emph{worst} input $\rho$ in some state set $\mathcal{S}$ to deteriorate Bob's testing.
There are two settings as described below.

\ul{\textbf{Setting I}: Bob is informed on which input state $\rho$ is used.}
Whenever Alice uses a state $\rho$ as input, Bob has to be \emph{informed} the identity of $\rho$ (e.g., by some third party Charlie), but Bob does not know which underlying channel  $\mathcal{N}_1$ or $\mathcal{N}_2$ is.
In other words, Bob can construct a test depending on $\rho$, and Alice may calculate Bob's distinguishability \emph{apriori} and choose the worst possible $\rho$ accordingly.

\ul{\textbf{Setting II}: Bob is NOT informed on which input state $\rho$ is used.}
In this setting, Bob has no ideas of which state Alice would use. Hence, his test cannot depend on $\rho$ and shall accommodate the worst output.

Unless Alice's choice of inputs $\mathcal{S}$ is a singleton (which reduces to the usual binary hypothesis testing of states), the distinguishability of Setting I should be greater than that of Setting II from the operational perspective because of Bob's prior information.
Indeed, our first result shows that, when Alice's choice is restricted to i.i.d.~states $\rho^{\otimes n}$ as input to i.i.d.~channel $\mathcal{N}_1^{\otimes n}$ or $\mathcal{N}_2^{\otimes n}$, Stein's exponent of {\bf Setting I (Bob informed)} is given by the 
\emph{infimum channel divergence}:
\begin{align} \label{eq:channel_divergence_1}
D(\N_1\|\N_2):= \inf_{\rho}D(\N_1(\rho)\|\N_2(\rho))\ ,
\end{align}
and Stein's exponent of {\bf Setting II (Bob non-informed) } is given by
\begin{align} \label{eq:D_inf_1}
D^{\text{inf}}(\N_1||\N_2):= \inf_{\rho,\sigma}D(\N_1(\rho)\|\N_2(\sigma))\ ,
\end{align}
where the infimum is over all pairs of input states (see Theorem~\ref{thm:iid}).
As will be shown in Example~\ref{exam:1} later, there are channels $\mathcal{N}_1, \mathcal{N}_2$ such that $D^{\text{inf}}(\N_1\|\N_2) = 0$ while $D(\N_1\|\N_2)$ could be strictly positive.

When Alice is allowed to choose general correlated inputs to i.i.d.~channels, Bob in Setting II (non-informed) is essentially distinguishing the collection $\{\N_1^{\ten n}(R)\}_R$ from $\{\N_2^{\ten n}(S)\}_S$, where for each $n$, the index is over all pairs of general input states $(R,S)$ on the $n$-fold quantum system.
This scenario was studied by Fang, Fawzi and Fawzi in \cite{fang2025adversarial} under the name \emph{adversarial channel discrimination}, and they showed Stein's exponent is given by the regularization of \eqref{eq:D_inf_1}:
\begin{align} \label{eq:D_inf_reg}
D^{\text{inf},\infty}(\N_1\|\N_2):=\lim_{n\to \infty}\frac{1}{n} D^{\text{inf}}(\N_1^{\ten n}\|\N_2^{\ten n}).
\end{align}
On the other hand, we show that Stein's exponent in the Setting I (informed) is the regularization of \eqref{eq:channel_divergence_1}:
\begin{align} \label{eq:channel_divergence_reg}
D^{\infty}(\N_1\|\N_2):=\lim_{n\to \infty}\frac{1}{n} D(\N_1^{\ten n}\|\N_2^{\ten n}).
\end{align}
Moreover, we find that the Stein exponents for Settings I and II with general inputs are exactly \emph{the same} for all pairs of channels (see Theorem~\ref{thm:adversarial}):
\begin{align} \label{eq:equal}
D^{\text{inf},\infty}(\N_1\|\N_2) 
= D^{\infty}(\N_1\|\N_2).
\end{align}
This means that, surprisingly, the Bob's prior information in Setting~I does not make a difference to the Stein's exponent in the asymptotic limits.
See Table~\ref{table:QQ} for summary.

In adversarial channel discrimination, regularization in \eqref{eq:D_inf_reg} generally required \cite{FFF}.
For a special case that the channels are entanglement-breaking (EB) with rank-one projective measurements of the form:
\begin{align*}
\N_1(\rho)\!=\!\!\sum_{x\in {\cal X}}
\langle v_{x}|\rho|v_{x} \rangle \rho_{1,x}, \; \N_2(\rho)\!=\!\!\sum_{x\in {\cal X}}
\langle u_{x}|\rho|u_{x} \rangle \rho_{2,x}, 
\end{align*}
Stein's exponent with general inputs can be attained by that with i.i.d.~inputs, and hence we obtain a single-letter expression (see Theorem~\ref{thm:EB}):
\begin{align} 
D^{\inf,\infty}(\N_1\|\N_2)
= D^{\inf}(\N_1\|\N_2) =\inf_{p,p'}D\left(\sum\nolimits_{x}p(x)\rho_{1,x}\|\sum\nolimits_{x}p'(x) \rho_{2,x}\right), \label{eq:EB_result}
\end{align}
where $p,p'$ are probability distributions on $\mathcal{X}$.

\begin{table}[t]
	\centering
	\resizebox{0.7\columnwidth}{!}{
		\begin{tabular}{@{}ccc@{}}
        
        \toprule
        \multirow{2}{*}{QQ channels $\mathcal{N}_i$} & \textbf{Setting I} & \textbf{Setting II}
        \\
        & (Bob informed) & (Bob non-informed)
        \\
        \hline

        general inputs & \multicolumn{2}{c}{\multirow{2}{*}{$ \displaystyle  D^{\infty}(\N_1\|\N_2) = D^{\text{inf}, \infty}(\N_1\|\N_2) $}}

        \\
        
        (Theorem~\ref{thm:adversarial}) & & 

        \\

        \arrayrulecolor{black!25}
        \cmidrule{1-2}
        
        i.i.d.~inputs & \multirow{2}{*}{$\displaystyle D(\N_1\|\N_2)$} &  
        \multirow{-2.6}{*}{{\color{black!35}\scriptsize (equal for EB channels)}}

        \\

        (Theorem~\ref{thm:iid}) &   & 
        \multirow{-2}{*}{$\displaystyle D^{\text{inf}}(\N_1\|\N_2)$}

        \\
            
        \arrayrulecolor{black}\bottomrule
        
        \end{tabular}
	}
	\caption{
    \textnormal{
	Summary of the Stein exponents for adversarial hypothesis testing of quantum-quantum (QQ) channels.
    The quantities $D(\mathcal{N}_1\|\mathcal{N}_2)$, $D^{\text{inf}}(\mathcal{N}_1\|\mathcal{N}_2)$, $D^{\infty}(\mathcal{N}_1\|\mathcal{N}_2)$, and $D^{\text{inf},\infty}(\mathcal{N}_1\|\mathcal{N}_2)$ are given in \eqref{eq:channel_divergence_1}, \eqref{eq:D_inf_1}, \eqref{eq:channel_divergence_reg}, \eqref{eq:D_inf_reg}, respectively.
    Note that for entanglement-breaking channels with rank one projective measurements, $D^{\infty}(\N_1\|\N_2) = D^{\text{inf}, \infty}(\N_1\|\N_2) = D^{\text{inf}}(\N_1\|\N_2)$  (Theorem~\ref{thm:EB}), which could be different from $D(\N_1\|\N_2)$.
    }
	}	\label{table:QQ}	
\end{table}

Next, we move on to the scenario of classical-quantum (CQ) channels: $W_i: x\mapsto \rho_{i,x}$, for which Alice chooses a classical symbol $x\in\mathcal{X}$ at input and Bob receives a quantum state $\rho_{i,x}$ at output of channel $W_i$.
In \textbf{Setting I}, {\bf we assume Bob is informed about the realization of $x\in\mathcal{X}$ which has been chosen} and then Alice may randomize the input $x$ by some probability distribution $p$ on $\mathcal{X}$. In the adversarial setting, such assumption is natural for CQ channels.
Hence, the effective channel on Bob side's is
\[\tilde{W}_i(p) \mapsto \sum\nolimits_{x\in\mathcal{X}} p(x)\,|x\rangle\!\langle x|\otimes \rho_{i,x} \ ,\]

\noindent
where the outputs for discrimination are CQ states.
We show that Stein's exponents for both the general-input case (i.e., Alice can choose any probability distributions $p^n$ on $\mathcal{X}^n$) and the i.i.d.-input case (i.e.,~Alice chooses $p^{\otimes n}$) are given by the single-letter channel divergence:
\begin{align} \label{eq:CQ_informed}
D(\tilde{W}_1 \| \tilde{W}_2) =
\inf_{x\in\mathcal{X}} D(\rho_{1,x}\|\rho_{2,x}).
\end{align}

In \textbf{Setting II}, where the input signal $x$ is not visible to Bob, we show that Stein's exponents for both the general-input and the i.i.d.-input cases are given by 
\begin{align}
D^{\inf }(W_1\|W_2)=\inf_{p,p'}D\left(\sum\nolimits_{x}p(x)\rho_{1,x}\|\sum\nolimits_{x}p'(x) \rho_{2,x}\right).
\end{align}
Indeed, such a result is an immediate consequence of \eqref{eq:EB_result} for EB channels with rank one projective measurements.
However, we emphasize that \textbf{Setting I} (informed) manifests a surprising distinction---\eqref{eq:equal} no longer always holds for CQ channels.
Further, there are CQ channels (see Example~\ref{exam:1}) such that
\begin{align} \notag
D^{\infty}(\tilde{W}_1 \| \tilde{W}_2)=D(\tilde{W}_1 \| \tilde{W}_2)
> D^{\inf}(W_1||W_2)=D^{\inf,\infty }(W_1||W_2)
\end{align}
which means that informed Bob in Setting I is much powerful than being non-informed because of the visible signal $x$ and the CQ structure.
This adversarial hypothesis testing problem also demonstrates a striking phenomenon that \emph{results of EB channels do not always cover that of CQ channels}, because coherence at the input system yields inherent randomness that could make a fundamental difference.
We refer readers to Table~\ref{table:CQ} for summary of CQ channels.

\begin{table}[ht]
	\centering
	\resizebox{0.7\columnwidth}{!}{
		\begin{tabular}{@{}ccc@{}}

        \toprule
        CQ channels  & \;\;\;\;\;\textbf{Setting I} & \;\textbf{Setting II}
        \\
        $\displaystyle W_i: x\mapsto \rho_{i,x}$ & \;\;\;\;\;(Bob informed) & (Bob non-informed)
        \\

        \hline

        \multirow{2}{*}{general inputs} & \multicolumn{2}{c}{\multirow{4.7}{*}{$ \displaystyle  \inf_{x\in\mathcal{X}} D(\rho_{1,x}\|\rho_{2,x}) {\;\;\color{black!30}\geq\;} \inf_{p,p'}D\left(W_1(p) \| W_2(p')\right) $}}
        \\
        
         & & 

        \\

        \arrayrulecolor{black!25}\cmidrule{1-1}
        
        \multirow{2}{*}{i.i.d.~inputs} 
        &

        \\

         & (Theorem~\ref{thm:cqinform})  & (Theorem~\ref{thm:cqnoninform})

        \\
            
        \arrayrulecolor{black}\bottomrule
        
        \end{tabular}
	}
	\caption{
    \textnormal{
	Summary of the Stein exponents for adversarial hypothesis testing of classical-quantum (CQ) channels.
    Here, $W_i(p) := \sum_{x\in\mathcal{X}} p(x) \rho_{i,x}$.
    Note that the first row is `$\geq$' and the quantities in the first column are equal, as opposed to Table~\ref{table:QQ}.
    }
	}	\label{table:CQ}	
\end{table}

\begin{table}[ht]
	\centering
	\resizebox{0.7\columnwidth}{!}{
		\begin{tabular}{@{}ccc@{}}

        \toprule
        \textbf{Setting I}  & \multirow{2}{*}{\;\;\;\;\;\;\;CQ Channels}  & \multirow{2}{*}{\;\;\;EB Channels}
        \\
        (Informed) &  & 
        \\

        \hline

        \multirow{2}{*}{general inputs} & \multicolumn{2}{c}{\multirow{2.4}{*}{$ \displaystyle  \;\;\inf_{x\in\mathcal{X}} D(\rho_{1,x}\|\rho_{2,x}) {\;\;\color{black!30}\geq\;\;}  \inf_{p,p'}D\left(W_1(p) \| W_2(p')\right) $}}
        \\
        
         & & 

        \\

        \arrayrulecolor{black!25}\midrule
        
        \multirow{2}{*}{i.i.d.~inputs} 
        &
        \multicolumn{2}{c}{\multirow{2}{*}{$ \displaystyle  \inf_{x\in\mathcal{X}} D(\rho_{1,x}\|\rho_{2,x}) {\;\;\color{black!30}\geq\;\;} \inf_{p}D\left(W_1(p) \| W_2(p)\right) $}}

        \\

         &   &  

        \\
            
        \arrayrulecolor{black}\bottomrule
        
        \end{tabular}
	}
	\caption{
    \textnormal{
	Distinction between classical-quantum (CQ) channels $W_i:x\mapsto \rho_{i,x}$ and the corresponding entanglement-breaking (EB) channels 
    $\N_i: \sigma \mapsto \sum_{x\in\mathcal{X}} \langle x | \sigma | x \rangle \rho_{i,x}$
    in Setting I (informed).
    Here, $W_i(p) := \sum_{x\in\mathcal{X}} p(x) \rho_{i,x}$.
    }
	}	\label{table:CQ_vs_EB}	
\end{table}

This paper is organized as follows.
Section~\ref{sec:QQ} discusses preliminary definitions and adversarial hypothesis testing for quantum-quantum channels.
Section~\ref{sec:CQ} does for classical-quantum channels.
We conclude this paper in Section~\ref{sec:discussions}.

{\bf Acknowledgement. }  The authors are grateful to Hayata Yamasaki for helpful discussions on adversarial hypothesis testing for CQ channels.
M.H. was supported in part by the Guangdong Provincial Quantum Science Strategic Initiative (Grant No.~GDZX2505003), the General R\&D Projects of 1+1+1 CUHK-CUHK(SZ)-GDST Joint Collaboration Fund (Grant No.~GRDP2025-022),
and the Shenzhen International Quantum Academy (Grant No. SIQA2025KFKT07).
H.C.~is supported by NSTC 113-2119-M-001-009, No.~NSTC 113-2628-E-002-029, NSTC 114-2124-M-002-003, NTU-113V1904-5, No.~NTU-114L895005, and No.~NTU-114L900702. L.G.~is partially supported by the National Natural Science Fundation of China (grant No.~12401163), by the Hubei Provincial International Collaboration  (Project No. 2025EHA041), and Hubei Provincial Innovationa research group (Project No. 2025AFA044).

\section{Adversarial Hypothesis Testing for QQ channels} \label{sec:QQ}
\subsection{Preliminary and Results}
Let $\rho$ and $\sigma$ be two quantum states. For binary hypothesis testing, one wants to distinguish $\rho$ from $\sigma$ via implementing some test operator $0\le T\le I$.  The type I error $\alpha$ and type II error $\beta$ are defined as
\[ \alpha=\tr((I-T)\rho))\ , \beta=\tr(T\sigma)\pl. \]
In the regime of the Stein's lemma, we are interested in minimizing the type II error given the Type I error is small. For $\varepsilon\in (0,1)$, define
\begin{align*}\beta_\varepsilon(\rho,\sigma):=\inf \{\tr(T\sigma)\ |\ 0\le T\le I , \tr((I-T)\rho)\le \varepsilon\}\ , 
\end{align*}
and the hypothesis testing relative entropy is defined as 
\[D_h^\varepsilon(\rho||\sigma):=-\log\beta_\varepsilon(\rho||\sigma)\ .\]
The Stein's Lemma \cite{HP91, NO00} states that in the i.i.d. setting,  the type II error decays exponentially, and the exponent is given by
\begin{align} \lim_{n\to \infty}-\frac{1}{n}\log\beta_\varepsilon(\rho^{\otimes n},\sigma^{\otimes n})=\lim_{n\to \infty}\frac{1}{n}D_h^\varepsilon(\rho^{\otimes n}||\sigma^{\otimes n}) =D(\rho||\sigma)\pl.\label{eq:stein}\end{align}
where $D(\rho||\sigma)=\tr(\rho\log \rho -\rho\log \sigma)$ is the standard relative entropy.

Now let us consider the binary hypothesis testing for quantum channels.
Let $\mathcal{N}_1,\mathcal{N}_2:\mathcal{B}(\mathcal{H}_A)\to \mathcal{B}(\mathcal{H}_B)$ be two quantum channels. The setting of the problem is: the sender Alice sends an input state $\rho$,
through either the channel $\mathcal{N}_1$ (the  null hypothesis), or the channel $\mathcal{N}_2$
(the alternative hypothesis). Then the receiver Bob will implement some test $T$ on the output, which will either be $\mathcal{N}_1(\rho)$ or $\mathcal{N}_2(\rho)$, to distinguish the case $\mathcal{N}_1$ being used from that $\mathcal{N}_2$ is used. The type I error and Type II error are defined as
\[ \alpha=\tr((I-T)(\mathcal{N}_1(\rho)))\ , \beta=\tr(T\mathcal{N}_2(\rho))\pl, \]
In this work, we consider Alice is  {\bf adversarial} in the sense that {\bf she may choose the \emph{worst} input $\rho$ to maximize the error probability $\beta_\varepsilon$}.
There are two settings: \\
{\bf Setting I}: \underline{Bob is informed on which input state $\rho$ is used;}\\
{\bf Setting II}: \underline{Bob is NOT informed on which input state $\rho$ is used.}\\
In {\bf Setting I}, when the input state $\rho$ is informed to Bob, Bob can choose the test $T$ depending on Alice's choice of $\rho$, the error probability is
\[ \beta_{\varepsilon}(\N_1,\N_2):=\max_{\rho}\inf_{0\le T\le I} \{\tr(T\N_2(\rho))\ |\   \tr((I-T)\N_1(\rho))\le \eps\} \]
We define the hypothesis testing relative entropy for two channels
\[ D_h^\varepsilon(\N_1\|\N_2):=-\log \beta_{\varepsilon}(\N_1,\N_2).\]
In {\bf Setting II}, when the input state $\rho$ is not informed to Bob, Bob uses
a test $T$ independent of the input $\rho$. In this case, Bob is actually discriminate between the two states sets $\{\N_1(\rho)\}_{\rho\in \mathcal{D}(\mathcal{H}_A)}$ and $\{\N_2(\sigma) \}_{\sigma\in \mathcal{D}(\mathcal{H}_A)} $, whose error probability is
\[ \beta_{\varepsilon}(\{\N_1(\sigma)\}_{\rho},\{\N_2(\sigma) \}_{\sigma})=\inf_{0\le T\le I} \left\{ \max_{\sigma \in \mathcal{D}(\mathcal{H}_A) }\tr(T\N_2(\rho))\ |\  \tr((I-T)\N_1(\rho))\le \eps, \forall \rho \in \mathcal{D}(\mathcal{H}_A) \right\} \ ,\]
and we denote the hypothesis testing relative entropy for two sets of states as follow
\[ D_h^\varepsilon(\{\N_1(\rho)\}_{\rho}||\{\N_2(\sigma) \}_{\sigma}):=-\log  \beta_{\varepsilon}(\{\N_1(\rho)\}_{\rho},\{\N_2(\sigma) \}_{\sigma})\ .\]
In general, $D_h^\varepsilon(\{\N_1(\rho)\}_{\rho}||\{\N_2(\sigma) \}_{\sigma})\le D_h^\varepsilon(\N_1\|\N_2)$.

Our first theorem is the Stein exponenet when Alice uses only i.i.d. inputs $\rho^{\otimes n}$. 
\begin{theorem}[Adversarial Stein's exponent with i.i.d.~inputs]\label{thm:iid}For any two quantum channels $\N_1, \N_2$ and $\varepsilon\in(0,1)$
\begin{align*}
&\lim_{n\to \infty}\frac{1}{n} \inf_{\rho} D_h^\varepsilon(\N_1^{\ten n}(\rho^{\ten n})\|\N_2^{\ten n}(\rho^{\ten n}))=D(\N_1\|\N_2),
\\ &\lim_{n\to \infty}\frac{1}{n} D_h^\varepsilon(\{\N_1^{\ten n}(\rho^{\ten n})\}_\rho \|\{\N_2^{\ten n}(\sigma^{\ten n})\}_\sigma)=D^{\textnormal{inf}}(\N_1\|\N_2).
\end{align*}
\end{theorem}
Recall the definition of two channel divergences
\begin{align} \label{eq:DD}
D(\N_1\|\N_2):=\inf_{\rho}D(\N_1(\rho)\|\N_2(\rho)) \ ,\  
D^{\text{inf}}(\N_1\|\N_2):= \inf_{\rho,\sigma}D(\N_1(\rho)\|\N_2(\sigma))\ .
\end{align}
It is clear from the definition that $D(\N_1\|\N_2)\ge D^{\text{inf}}(\N_1\|\N_2)$, which matches the operational perspective that Bob has prior information in {\bf Setting I}. 

We note that the one-shot quantity $D(\N_1\|\N_2)$ and $D^{\inf}(\N_1\|\N_2)$ can be quite different as illustrated in Example~\ref{exam:1} below.
\begin{exam}\label{exam:1}
Let $\mathcal{H}_A=\mathbb{C}^2$ and  $\{|0\rangle,|1\rangle\}$ be the standard basis. 
Consider the channels
\begin{align*}
\N_1(\rho)&:= \langle 0|\rho|0\rangle |0\rangle \langle 0|
+\frac{1}{2}\langle 1|\rho|1\rangle
 \mathbbm{1}, \\
\N_2(\rho)&:= \frac{1}{2}\langle 0|\rho|0\rangle 
  \mathbbm{1}
+\langle 1|\rho|1\rangle |1\rangle \langle 1|,
\end{align*}
where $\mathbbm{1}$ is the $2\times 2$ identity operator.
In this case,
\[D(\N_1\|\N_2 )=\inf_{\rho\in {\cal D}(\mathcal{H}_A)}
D(\N_1(\rho)\|\N_2(\rho))> 0\ , \ D^{\inf }(\N_1||\N_2 )=\inf_{\rho,\sigma\in {\cal D}(\mathcal{H}_A)}
D(\N_1(\rho)\|\N_2(\sigma))=0.\]
\end{exam}

Alternatively, it is also natural to consider that Alice can use joint inputs for $n$-copy of the channels,
Thus, in the $n$-shot settings, we are interested 
\begin{align*}
& \beta_{\varepsilon}(\N_1^{\ten n},\N_2^{\ten n})= \max_{R} \inf_{0\le T\le I}\{\tr(T\N_2^{\ten n}(R))\ |\ \tr((I-T)\N_1^{\ten n}(R))\le \eps\},
\\ & \beta_{\varepsilon}(\{\N_1^{\ten n}(R)\}_R,\{\N_2^{\ten n}(S)\}_S)= \inf_{0\le T\le I} \left\{ \max_{S}\tr(T\N_2^{\ten n}(S))\ |\  \tr((I-T)\N_1^{\ten n}(R))\le \eps, \text{ for any } R \right\}
\end{align*}
Here and in the following, we use the capital letter $R,S \in \mathcal{D}(\mathcal{H}_A^{\otimes n})$ to represent the joint input states for the $n$-shot channels $\N_1^{\ten n}$ and $\N_2^{\ten n}$. The second quantity  was studied by Fang, Fawzi and Fawzi in \cite{fang2025adversarial} under the name \emph{adversarial channel discrimination}.
They proved the following Stein's exponent
\[\lim_{n\to \infty}\frac{1}{n} D_h^\varepsilon(\{\N_1^{\ten n}(R)\}_R\|\{\N_2^{\ten n}(S)\}_S)=\lim_{n\to \infty}\frac{1}{n} \inf_{R,S}D( \N_1^{\ten n}(R)\|\N_2^{\ten n}(S)) ,\]
where for each $n$, the infimum is over all pairs of $n$-fold input states $R,S\in \mathcal{D}(\mathcal{H}_A^{\ten n})$. The Stein exponent in the right hand side is the regularization of $D^{inf}$ divergence
\[ D^{\text{inf},\infty}(\N_1\|\N_2):=\lim_{n\to \infty}\frac{1}{n} D^{\text{inf}}(\N_1^{\ten n}\|\N_2^{\ten n}) \  .\]
 In the same spirit, we introduce the regularization for channel divergence $D$ as follows
\[ D^{\infty}(\N_1\|\N_2):=\lim_{n\to \infty}\frac{1}{n} D(\N_1^{\ten n}\|\N_2^{\ten n}) .\]

Surprisingly, we find that when using general inputs, the {\bf Setting I (informed)} and {\bf Setting II (non-informed)} share the same Steins exponent. 
\begin{theorem}[Adversarial Stein's exponent with general inputs]\label{thm:adversarial}Let $\N_1$ and $\N_2$ be two quantum channels. For any $\varepsilon\in(0,1)$
\begin{align*}
\lim_{n\to \infty}\frac{1}{n} D_h^\varepsilon(\N_1^{\ten n}\|\N_2^{\ten n})=\lim_{n\to \infty}\frac{1}{n} D_h^\varepsilon(\{\N_1^{\ten n}(R)\}_R\|\{\N_2^{\ten n}(S)\}_S)=D^{\textnormal{inf},\infty}(\N_1\|\N_2)=D^{\infty}(\N_1\|\N_2)\ .
\end{align*}
\end{theorem}
The above theorem implies that, for the adversarial hypothesis testing using general inputs, whether the inputs are informed to Bob or not does not change the Stein's exponent in the asymptotic limit. In particular, it yields an interesting observation that the channel divergences $D( \cdot\|\cdot )$ and $D^{\text{inf}}( \cdot\|\cdot )$ share the same regularization. 
\[ D^{\text{inf},\infty}(\N_1\|\N_2)=D^{\infty}(\N_1\|\N_2).\]

\subsection{Proof of Theorem \ref{thm:iid}: i.i.d.~inputs}

Let
$\mathcal{S}_1,\mathcal{S}_2\subset \mathcal{B}(\cH)_+$ be two sets of positive operators.
Define the discrimination error
\begin{align} \label{eq:binary_group}
\varepsilon(\mathcal{S}_1 \| \mathcal{S}_2)
:= 
\inf_{0\leq T\leq I} \left\{
\sup_{\rho \in \mathcal{S}_1} \left\{\tr \left[\rho (I-T)\right]\right\} +
\max_{\sigma \in \mathcal{S}_2} \left\{\tr \left[\sigma T\right]\right\} \right\}.
\end{align}
Also, recall that the Petz-R\'enyi relatvie entropy 
\[D_{1-s}(\rho\|\sigma)=\frac{1}{s-1}\log tr\left[ \rho^{s} \sigma^{1-s} \right], s\in (0,1)\]
We will use the following lemma essentially showed by Audenaert and Mosonyi \cite{AM}.
\begin{lemma}[{\cite{AM}}] \label{AM}
For two finite sets of states
$\{\rho_1,\ldots,\rho_{r_1}\}$
and $\{\sigma_1,\ldots,\sigma_{r_2}\}$,
\begin{align}
	\varepsilon(\{\rho_1,\ldots,\rho_{r_1}\} \| \{\sigma_1,\ldots,\sigma_{r_2}\})	
	\leq
\sum_{1\le j\le r_1, \, 1\le {j'}\le r_2}  \min_{s\in[0,1]} \sqrt{ 2 \emph{tr}\left[ \rho_j^{1-s} \sigma_{j'}^s \right] }. \label{BAG1}
\end{align} 
\end{lemma}
\begin{proof}We provide a proof for completeness.  We first upper bound the maximum by summation in \eqref{eq:binary_group}:
\begin{align}
\varepsilon(\{\rho_1,\ldots,\rho_{r_1}\} \| \{\sigma_1,\ldots,\sigma_{r_2}\})
&\leq \inf_{0\leq T\leq I} \left\{
\sum_{j\in[r_1]} \tr \left[\rho_j (I-T)\right] +
\sum_{j'\in[r_2]} \tr \left[\sigma_{j'} T\right] \right\}
\\
&\overset{\textnormal{(a)}}{\leq} F\left( \sum\nolimits_{j\in[r_1]} \rho_j , \sum\nolimits_{j'\in[r_2]} \sigma_{j'}\right)
\\
&\overset{\textnormal{(b)}}{\leq} \sum_{j\in[r_1], \, {j'}\in[r_2]} F\left( \rho_j, \sigma_{j'} \right)
\\
&\overset{\textnormal{(c)}}{\leq} \sum_{j\in[r_1], \, {j'}\in[r_2]} \sqrt{ 2 - \left\| \rho_j - \sigma_{j'} \right\|_1   }
\\
&\overset{\textnormal{(d)}}{\leq}  \sum_{j\in[r_1], \, {j'}\in[r_2]}  \min_{s\in[0,1]} \sqrt{ 2 \tr\left[ \rho_j^{1-s} \sigma_{j'}^s \right] }.
\end{align} 
Here, (a) follows from the Fuchs--van de Graaf inequality \cite{FG, Audenaert}, and the fidelity for positive semi-definite operators $A$ and $B$ is defined as
\begin{align}
	F(A,B) := \left\| \sqrt{A} \sqrt{B} \right\|_1
	= \tr \left[ \left( A^{1/2} B A^{1/2} \right)^{1/2} \right].
\end{align}
The inequality (b) follows from the subadditivity of fidelity \cite[Lemma 4.9]{AM}.
The inequality (c) follows from the Fuchs--van de Graaf inequality, again (see e.g., \cite[Lemma 2.4]{AM}).
The last inequality (d) follows from the quantum Chernoff bound \cite{ACMMABV}.
\end{proof}

\begin{proof}[Proof of Theorem \ref{thm:iid}]
The Stein's lemma \eqref{eq:stein} implies
\begin{align}
\lim_{n\to \infty}\frac{1}{n}\inf_{\rho\in {\cal D}({ H}_A)}
D^h_\epsilon(\mathcal{N}_1^{\otimes n}(\rho^{\otimes n})\|\mathcal{N}_2^{\otimes n}(\rho^{\otimes n}))
\le 
\inf_{\rho\in {\cal D}({H}_A)}
D(\mathcal{N}_1(\rho)\|\mathcal{N}_2(\rho)) 
\label{EQ4},
\end{align}
and
\begin{align}
\lim_{n\to \infty}\frac{1}{n}
D^h_\epsilon(
\{\mathcal{N}_1^{\otimes n}(\rho^{\otimes n})\}_{\rho\in {\cal D}({ H}_A)}\|\{\mathcal{N}_2^{\otimes n}(\sigma^{\otimes n})\}_{\sigma\in {\cal D}({H}_A)})
\le 
\inf_{\rho,\sigma\in {\cal D}({ H}_A)}
D(\mathcal{N}_1(\rho)\|\mathcal{N}_2(\sigma)) 
\label{EQ4I}.
\end{align}
We show the opposite inequality 
\begin{align}
\lim_{n\to \infty}\frac{1}{n}
D^h_\epsilon(
\{\mathcal{N}_1^{\otimes n}(\rho^{\otimes n})\}_{\rho\in {\cal D}({H}_A)}\|\{\mathcal{N}_2^{\otimes n}(\sigma^{\otimes n})\}_{\sigma\in {\cal D}({H}_A)})
\ge 
\inf_{\rho,\sigma\in {\cal D}({H}_A)}
D(\mathcal{N}_1(\rho)\|\mathcal{N}_2(\sigma)) 
\label{EQ43}.
\end{align}
and the opposite inequality  to \eqref{EQ4} can be derived similarly.
We use the method in \cite[Assumption 1]{DWH}. 
For any $\epsilon>0$, 
there exists a finite set ${\cal S} \subseteq \mathcal{D}({H}_A)$
such that for any element $ \sigma \in {\cal D}({H}_A)$,
there exists an element $\rho[\sigma]\in {\cal D} $
such that 
\begin{align}
\sigma \le e^{\epsilon}\rho[\sigma].\label{CBN}
\end{align}
Indeed, for every positive state $\rho \in {\cal D}({H}_A)$, we define the open set
$U[\epsilon,\rho]$ as
\begin{align}
U[\epsilon,\rho]:= \{\sigma \in {\cal D}({H}_A)
| \sigma < e^{\epsilon}\rho \}.
\end{align}
Since ${\cal D}({ H}_A)$ is a compact set,
there exists a finite subset ${\cal D}\subset {\mathcal{D}}({H}_A)$ such that
$\cup_{\rho \in {\cal D}}U[\epsilon,\rho]
= {\cal D}({ H}_A)$.

Now choose $r$ as
\begin{align}
r < \inf_{\rho,\sigma \in {\cal D}({H}_A)}
D( \mathcal{N}_1(\rho)\| \mathcal{N}_2(\sigma)).\label{ZNO}
\end{align}
Then we  choose $0<s<1$ such that
\begin{align}
r < \inf_{\rho,\sigma \in {\cal D}({H}_A)}
D_{1-s}( \mathcal{N}_1(\rho)\| \mathcal{N}_2(\sigma)), 
\end{align}
and $\epsilon >0$ such that
\begin{align}
\epsilon< \frac{(1-s)}{2}\left(\inf_{\rho,\sigma \in {\cal D}({H}_A)}
D_{1-s}( \mathcal{N}_1(\rho)\| \mathcal{N}_2(\sigma)) -r\right).
\end{align}
By the choice of the subset ${\cal S}\subset {\cal D}({H}_A)$ and $N=|{\cal S}|$,
 we have
\begin{align}
\epsilon< \frac{(1-s)}{2}\left(\inf_{\rho,\sigma \in {\cal S}}
D_{1-s}( \mathcal{N}_1(\rho)\| \mathcal{N}_2(\sigma)) -r\right).
\label{BLP}
\end{align}
By the condition Eq. \eqref{CBN}, for any state 
$\rho \in {\cal D}({H}_A)$,
there exists a state $\rho_0 \in {\cal S}$ such that
\begin{align}
\rho^{\otimes n}
\le e^{n\epsilon} \rho_0^{\otimes n},
\end{align}
which implies
\begin{align}
\N_i^{\otimes n}(\rho^{\otimes n})
\le e^{n\epsilon}\N_i^{\otimes n}(\rho_0^{\otimes n})\ ,\ \text{ for } i=1,2
\label{CBN2}
\end{align}
Thus, we have
\begin{align}
&\varepsilon(
\{\mathcal{N}_1^{\otimes n}(\rho^{\otimes n})\}_{\rho\in{\cal D}({H}_A)}\|\{e^{nr}\mathcal{N}_2^{\otimes n}(\sigma^{\otimes n})\}_{\sigma\in{\cal D}({H}_A)}) \\
\stackrel{(a)}{\le}&
\varepsilon(
\{e^{n\epsilon}\mathcal{N}_1^{\otimes n}(\rho^{\otimes n})\}_{\rho\in{\cal D}}\|
\{e^{nr}e^{n\epsilon} \mathcal{N}_2^{\otimes n}(\sigma^{\otimes n})\}_{\sigma\in{\cal D}}) \\
\stackrel{(b)}{\le}&
e^{n\epsilon} \cdot \sqrt{2} N^2
e^{-n\frac{(1-s)}{2} (
\inf_{\rho,\sigma \in {\cal D}}
D_{1-s}( \mathcal{N}_1(\rho)\| \mathcal{N}_2(\sigma)) -r)} \\
\stackrel{(c)}{\to} & 0,\label{ZNO2}
\end{align}
where 
$(a)$ follows from \eqref{CBN2},
$(b)$ follows from Lemma \ref{AM},
$(c)$ follows from \eqref{BLP}. Note that 
\begin{align*}
\lim_{n\to \infty}\varepsilon(
\{\mathcal{N}_1^{\otimes n}(\rho^{\otimes n})\}_{\rho\in{\cal D}({H}_A)}\|\{e^{nr}\mathcal{N}_2^{\otimes n}(\sigma^{\otimes n})\}_{\sigma\in{\cal D}({H}_A)})= 0 
\end{align*}
implies 
\begin{align}
\lim_{n\to \infty}\frac{1}{n}
D^h_\epsilon(
\{\mathcal{N}_1^{\otimes n}(\rho^{\otimes n})\}_{\rho\in {\cal D}({H}_A)}\|\{\mathcal{N}_2^{\otimes n}(\sigma^{\otimes n})\}_{\sigma\in {\cal D}({H}_A)})
\ge r
\label{EQ43}.
\end{align}
Since $r$ is an arbitrary real number satisfying \eqref{ZNO},
we obtain \eqref{EQ43}.
\end{proof}

\subsection{Proof of Theorem \ref{thm:adversarial}: general inputs}
We start with showing the regularized version of the two channel divergences coincides.   

\begin{lemma}\label{lemma:identity}
For two quantum channels $\N_1$ and $\N_2$,
\begin{align*}
D^{\emph{inf},\infty}(\N_1\|\N_2)=D^{\infty}(\N_1\|\N_2).
\end{align*}
Namely, we have 
\begin{align*}
\lim_{n\to \infty}\frac{1}{n} \inf_{R} D(\N_1^{\ten n}(R)\|\N_2^{\ten n}(R))= \lim_{n\to \infty}\frac{1}{n} \inf_{R,S} D(\N_1^{\ten n}(R)\|\N_2^{\ten n}(S)).
\end{align*}
\end{lemma}

\begin{proof}
Since
\begin{align}
\liminf_{n\to \infty}\frac{1}{n}\inf_{R}
D(\mathcal{N}_1^{\otimes n}(R)\|\mathcal{N}_2^{\otimes n}(R))
\ge
\lim_{n\to \infty}\frac{1}{n}
\inf_{R,S}
D(\mathcal{N}_1^{\otimes n}(R)\|\mathcal{N}_2^{\otimes n}(S)),
\label{EQ1}
\end{align}
it suffices to show the converse inequality 
\begin{align}
\limsup_{n\to \infty}\frac{1}{n}\inf_{R}
D(\mathcal{N}_1^{\otimes n}(R)\|\mathcal{N}_2^{\otimes n}(R))
\le
\lim_{n\to \infty}\frac{1}{n}
\inf_{R,S}
D(\mathcal{N}_1^{\otimes n}(R)\|\mathcal{N}_2^{\otimes n}(S)).
\label{EQ2A}
\end{align}
We show this by two steps.\\

{\bf Step 1.} First, we show that
\begin{align}
\limsup_{n\to \infty}\frac{1}{n}\inf_{R}
D(\mathcal{N}_1^{\otimes n}(R)\|\mathcal{N}_2^{\otimes n}(R))
\le
\inf_{\rho,\sigma}
D(\mathcal{N}_1(\rho)\|\mathcal{N}_2(\sigma)) =: D^{\inf}(\mathcal{N}_1\|\mathcal{N}_2)
\label{EQ2}.
\end{align}
Let $\rho,\sigma \in {\cal D}({H}_A)$ and $\epsilon>0$.
We choose the input state 
$(1-\epsilon)\rho^{\otimes n} +
\epsilon \sigma^{\otimes n} $. Since
\begin{align}
(1-\epsilon)\mathcal{N}_2(\rho)^{\otimes n} 
+\epsilon \mathcal{N}_2(\sigma)^{\otimes n}
\ge 
\epsilon \mathcal{N}_2(\sigma)^{\otimes n},
\end{align}
by operator monotonicity of $\log$ function,
\begin{align}
- \log( (1-\epsilon)\mathcal{N}_2(\rho)^{\otimes n} 
+\epsilon \mathcal{N}_2(\sigma)^{\otimes n})
\le 
- \log(\epsilon \mathcal{N}_2(\sigma)^{\otimes n}). \label{eq:mono}
\end{align}
 We have
\begin{align*}
&
D(\mathcal{N}_1^{\otimes n}((1-\epsilon)\rho^{\otimes n} +
\epsilon \sigma^{\otimes n})\|\mathcal{N}_2^{\otimes n}((1-\epsilon)\rho^{\otimes n} +
\epsilon \sigma^{\otimes n}))\\
=
&
D(
(1-\epsilon)\mathcal{N}_1(\rho)^{\otimes n} 
+\epsilon \mathcal{N}_1(\sigma)^{\otimes n})\|
(1-\epsilon)\mathcal{N}_2(\rho)^{\otimes n} 
+\epsilon \mathcal{N}_2(\sigma)^{\otimes n})
) \\
\le &
-H((1-\epsilon)\mathcal{N}_1(\rho)^{\otimes n} 
+\epsilon \mathcal{N}_1(\sigma)^{\otimes n})) -
\tr ((1-\epsilon)\mathcal{N}_1(\rho)^{\otimes n} 
+\epsilon \mathcal{N}_1(\sigma)^{\otimes n}))
 \log(\epsilon \mathcal{N}_2(\sigma)^{\otimes n})) \\
\le &
-(1-\epsilon)H(\mathcal{N}_1(\rho)^{\otimes n} )
-\epsilon H(\mathcal{N}_1(\sigma)^{\otimes n}) -
\tr ((1-\epsilon)\mathcal{N}_1(\rho)^{\otimes n} 
+\epsilon \mathcal{N}_1(\sigma)^{\otimes n}))
 \log(\epsilon \mathcal{N}_2(\sigma)^{\otimes n})) \\
=&
(1-\epsilon)D(
\mathcal{N}_1(\rho)^{\otimes n} \|
\mathcal{N}_2(\sigma)^{\otimes n})
+\epsilon D(\mathcal{N}_1(\sigma)^{\otimes n}
\|\mathcal{N}_2(\sigma)^{\otimes n})
-\log \epsilon  \\
=&
n (1-\epsilon)D(
\mathcal{N}_1(\rho) \|\mathcal{N}_2(\sigma))
+n \epsilon D(\mathcal{N}_1(\sigma)\|\mathcal{N}_2(\sigma))
-\log \epsilon .
\end{align*}
Here the first inequality above follows from the operator inequality \eqref{eq:mono}, and the second inequality follows from the concavity of von Neumann entropy $H(\rho)=-\tr(\rho \log \rho )$.
Thus, 
\begin{align}
&\limsup_{n\to \infty}\frac{1}{n}
D(\mathcal{N}_1^{\otimes n}((1-\epsilon)\rho^{\otimes n} +
\epsilon \sigma^{\otimes n})\|\mathcal{N}_2^{\otimes n}((1-\epsilon)\rho^{\otimes n} +
\epsilon \sigma^{\otimes n})) \\
\le&  (1-\epsilon)D(
\mathcal{N}_1(\rho) \|\mathcal{N}_2(\sigma))
+ \epsilon D(\mathcal{N}_1(\sigma)\|\mathcal{N}_2(\sigma))\ ,
\end{align}
which implies
\begin{align}
&\limsup_{n\to \infty}\frac{1}{n}\inf_{R}
D(\mathcal{N}_1^{\otimes n}(R)\|\mathcal{N}_2^{\otimes n}(R)) 
\le  (1-\epsilon)D(
\mathcal{N}_1(\rho) \|\mathcal{N}_2(\sigma))
+ \epsilon D(\mathcal{N}_1(\sigma)\|\mathcal{N}_2(\sigma))
\end{align}
Since $\epsilon>0$ is arbitrary, we have 
\begin{align}
&\limsup_{n\to \infty}\frac{1}{n}\inf_{R}
D(\mathcal{N}_1^{\otimes n}(R)\|\mathcal{N}_2^{\otimes n}(R)) 
\le  \inf_{\rho,\sigma}D(
\mathcal{N}_1(\rho) \|\mathcal{N}_2(\sigma)).
\end{align}
Taking infimum over all pairs $(\rho,\sigma)$ yields 
\eqref{EQ2}.\\

{\bf Step 2.} We show that for each $m\ge 1$,  
\begin{align}
\limsup_{n\to \infty}\frac{1}{n}\inf_{R}
D(\mathcal{N}_1^{\otimes n}(R)\|\mathcal{N}_2^{\otimes n}(R))
\le
\frac{1}{m}
\inf_{R,S}
D(\mathcal{N}_1^{\otimes m}(R)\|\mathcal{N}_2^{\otimes m}(S)).\label{EQ2B}
\end{align}
Eq.\eqref{EQ2} implies that for any $m\ge 1$,
\begin{align}
\limsup_{n'\to \infty}\frac{1}{n'm}\inf_{R}
D(\mathcal{N}_1^{\otimes n'm}(S)\|\mathcal{N}_2^{\otimes n'm}(S))
\le
\frac{1}{m}
\inf_{R,S}
D(\mathcal{N}_1^{\otimes m}(R)\|\mathcal{N}_2^{\otimes m}(S)).
\label{eq:EQ2B}
\end{align}

For $n=n' m+k$ with $0\le k <m$, by the data-processing inequality, we have
\begin{align}
	\frac{1}{n}\inf_{R}
	D(\mathcal{N}_1^{\otimes n}(R)\|\mathcal{N}_2^{\otimes n}(R))
	&=
	\frac{1}{n' m+k}
	\inf_{R}	D(\mathcal{N}_1^{\otimes (n'm+k)}(R)\|\mathcal{N}_2^{\otimes (n'm+k)}(R))
\nonumber	\\
	&\leq
	\frac{1}{n' m+k}
	\inf_{R}D(\mathcal{N}_1^{\otimes (n'm+m)}(R)\|\mathcal{N}_2^{\otimes (n'm+m)}(R))
\nonumber	\\
	&\le \frac{1}{n' m}
	\inf_{R}
	D(\mathcal{N}_1^{\otimes (n'+1)m}(R)\|\mathcal{N}_2^{\otimes (n'+1)m}(R)).\nonumber	\\
    &\le \frac{n'+1}{n'}\cdot \frac{1}{(n'+1)m}
	\inf_{R}
	D(\mathcal{N}_1^{\otimes (n'+1)m}(R)\|\mathcal{N}_2^{\otimes (n'+1)m}(R)).
    \label{EQ2C}
\end{align}
Here, with slight abuse of notation, we use $R$ for the infimum over the input state spaces of corresponding tensor systems. 
Thus, the combination of \eqref{eq:EQ2B} and \eqref{EQ2C}
yields \eqref{EQ2B}, which implies
\eqref{EQ2A} by taking the limit of $m\to \infty$.
\end{proof}

\begin{proof}[Proof of Theorem \ref{thm:adversarial}]
Note that by definition, 
\begin{align}
\lim_{n\to \infty}\frac{1}{n}
D_\epsilon^h(
\{\mathcal{N}_1^{\otimes n}(R)\}_R\|
\{\mathcal{N}_2^{\otimes n}(S)\}_S)
\le 
\lim_{n\to \infty}\frac{1}{n}\inf_{R}
D^h_\epsilon(\mathcal{N}_1^{\otimes n}(R)\|\mathcal{N}_2^{\otimes n}(R)).
\label{EQ3F}
\end{align}
Recall that the Stein exponent proved in \cite[Theorem 1]{fang2025adversarial}
\begin{align}\lim_{n\to \infty}\frac{1}{n} D_h^\varepsilon(\{\N_1^{\ten n}(R)\}_R\|\{\N_2^{\ten n}(S)\}_S)=\lim_{n\to \infty}\frac{1}{n} \inf_{R,S}D( \N_1^{\ten n}(R)\|\N_2^{\ten n}(S)) ,\label{eq:FFF} \end{align}
The combination of 
\eqref{eq:FFF} and \eqref{EQ3F}
implies 
\begin{align}
\lim_{n\to \infty}\frac{1}{n}
\inf_{R}
D^h_\epsilon(\mathcal{N}_1^{\otimes n}(R)\|\mathcal{N}_2^{\otimes n}(R))
\ge 
\lim_{m\to \infty}
\frac{1}{m}\inf_{R,S}
D(\mathcal{N}_1^{\otimes m}(R)\|\mathcal{N}_2^{\otimes m}(S))=:D^{\inf,\infty}(\mathcal{N}_1\|\mathcal{N}_2).
\label{EQ3G}
\end{align}
Thus, it suffices to show the opposite inequality
\begin{align}
\lim_{n\to \infty}\frac{1}{n}
\inf_{R}
D^h_\epsilon(\mathcal{N}_1^{\otimes n}(R)\|\mathcal{N}_2^{\otimes n}(R))
\le 
\lim_{m\to \infty}
\frac{1}{m}\inf_{R,S}
D(\mathcal{N}_1^{\otimes m}(R)\|\mathcal{N}_2^{\otimes m}(S))
\label{EQ3}.
\end{align}
Note that by the Stein's Lemma \eqref{eq:stein}, for each $m\ge 1$ and $R\in \mathcal{D}(H_A^{\ten m})$, 
\begin{align*}
D(\mathcal{N}_1^{\otimes m}(R)\|\mathcal{N}_2^{\otimes m}(R))=& \lim_{n\to \infty}\frac{1}{n}
D^h_\epsilon( \mathcal{N}_1^{\otimes m}(R)^{\ten n}\|\mathcal{N}_2^{\otimes nm}(R)^{\ten n}) \\
=& \limsup_{n\to \infty}\frac{1}{n}
D^h_\epsilon(\mathcal{N}_1^{\otimes nm}(R^{\ten n})\|\mathcal{N}_2^{\otimes nm}(R^{\ten n}))\\
\ge & \limsup_{n\to \infty}\frac{1}{n} \inf_{S\in \mathcal{D}(H_A^{\ten nm})}
D^h_\epsilon(\mathcal{N}_1^{\otimes nm}(S)\|\mathcal{N}_2^{\otimes nm}(S))
\end{align*}
Divide both sides by $m$, and taking the infimum over $R\in \mathcal{D}(H_A^{\ten m})$ yields,
\begin{align}
\limsup_{n\to \infty}\frac{1}{nm}\inf_{S}
D^h_\epsilon(\mathcal{N}_1^{\otimes nm}(S)\|\mathcal{N}_2^{\otimes nm}(S))
\le
\frac{1}{m}\inf_{R}
D(\mathcal{N}_1^{\otimes m}(R)\|\mathcal{N}_2^{\otimes m}(R)).
\label{EQ3A}
\end{align}
For $n=n' m+k$ with $0\le k <m$, we have,
\begin{align}
\frac{1}{n}\inf_{S}
D^h_\epsilon(\mathcal{N}_1^{\otimes n}(S)\|\mathcal{N}_2^{\otimes n}(S)) =&
\frac{1}{n'm+k}\inf_{S}
D^h_\epsilon(\mathcal{N}_1^{\otimes (n'm+k)}(S)\|\mathcal{N}_2^{\otimes (n'm+k)}(S)) \nonumber\\
\le&
\frac{1}{n'm} \inf_{S}
D^h_\epsilon(\mathcal{N}_1^{\otimes (n'm+m)}(S)\|\mathcal{N}_2^{\otimes (n'm+m)}(S)) \nonumber\\
= &
\frac{n'+1}{n'}\cdot \frac{1}{(n'+1)m} \inf_{S}
D^h_\epsilon(\mathcal{N}_1^{\otimes (n'+1)m}(S)\|\mathcal{N}_2^{\otimes (n'+1)m}(S)) ,\label{EQ3B}
\end{align}
where the inequality above is data processing.
The combination of \eqref{EQ3A} and \eqref{EQ3B}
implies
\begin{align}
\lim_{n\to \infty}\frac{1}{n}
\inf_{S}
D^h_\epsilon(\mathcal{N}_1^{\otimes n}(S)\|\mathcal{N}_2^{\otimes n}(S))
\le
\frac{1}{m}\inf_{R}
D(\mathcal{N}_1^{\otimes m}(R)\|\mathcal{N}_2^{\otimes m}(R)).
\label{EQ3D}
\end{align}
Taking the limit in \eqref{EQ3D} and using Lemma \ref{lemma:identity},
we have
\begin{align}
\lim_{n\to \infty}\frac{1}{n}
\inf_{S}
D^h_\epsilon(\mathcal{N}_1^{\otimes n}(S)\|\mathcal{N}_2^{\otimes n}(S))
\le &
\lim_{m\to \infty}
\frac{1}{m}\inf_{R}
D(\mathcal{N}_1^{\otimes m}(R)\|\mathcal{N}_2^{\otimes m}(R)) \nonumber\\
=&
\lim_{m\to \infty}
\frac{1}{m}\inf_{R,S}
D(\mathcal{N}_1^{\otimes m}(R)\|\mathcal{N}_2^{\otimes m}(S)),
\label{EQ3E}
\end{align}
which completes the proof.
\end{proof}

\subsection{EB channels with rank-one projective measurements}\label{sec:EB}
In this part, we assume that 
$\mathcal{N}_1$ and 
$\mathcal{N}_2$ have the following forms: 
\begin{align}
\N_1(\rho)=\sum_{x\in {\cal X}}
\langle v_{x}|\rho|v_{x} \rangle \rho_{1,x}, \ \ \N_2(\rho)=\sum_{x\in {\cal X}}
\langle u_{x}|\rho|u_{x} \rangle \rho_{2,x} \label{eq:form}
\end{align}
That is, they are EB channels with measurements onto an orthonormal bases $\{\ket{v_{x}}\}_{x\in \mathcal{X}}$ and $\{\ket{u_{x}}\}_{x\in \mathcal{X}}$
that are possibly different.

Let $|{\cal X}|=d$ and
let $T_d^n$ be the set of types on ${\cal X}$ with length $n$.
For $p \in T_d^n$, let $T_p$ be the set of elements of
${\cal X}^n$ whose type is $p$.
We denote the uniform distribution
on $T_p$ by $P_p$.
Any permutation invariant $P$ on ${\cal X}^n$
is written as 
$\sum_{p \in T_d^n} q(p)P_p$ for some probability distribution $q$ on the types $T_d^n$.
Also, for a sequence $\vec{x}=(x_1,\cdots,x_n)$, we define 
$\rho_{k,\vec{x}}=\rho_{k,x_1}\otimes \cdots \otimes \rho_{k,x_n}$ for $k=1,2$.

\begin{theorem} \label{thm:EB}
Let $\N_1$ and $\N_2$ be two EB channels in the form of \eqref{eq:form}. Then 
\begin{align}
\lim_{n\to \infty}\frac{1}{n}
D^h_\epsilon(
\{\mathcal{N}_1^{\otimes n}(R)\}_{R}\|\{\mathcal{N}_2^{\otimes n}(S)\}_{S})=\inf_{p,p'} D\left(\sum_{x}p(x)\rho_{1,x}\|\sum_{x}p'(x) \rho_{2,x}\right)
\label{XH1}
\end{align}
where  the infimum is over all distributions $p,p'$ on $\mathcal{X}$. 
As a consequence, we have   
\begin{align}
&\lim_{n\to \infty}\frac{1}{n}
D_\epsilon^h(
\{\mathcal{N}_1^{\otimes n}(R)\}_R\|
\{\mathcal{N}_2^{\otimes n}(S)\}_S)
=D^{\inf,\infty}(\N_1\|\N_2)
\\
=&\lim_{n\to \infty}\frac{1}{n}
D^h_\epsilon(
\{\mathcal{N}_1^{\otimes n}(\rho^{\otimes n})\}_{\rho}\|\{\mathcal{N}_2^{\otimes n}(\sigma^{\otimes n})\}_{\sigma}) 
=D^{\inf}(\N_1\|\N_2)
\label{XH9}
\end{align}
\end{theorem}
\if
Also, we use
\begin{align}
{H}_B^{\otimes n}=
\bigoplus_{\lambda \in\Lambda_n}
{\cal U}_{\lambda}\otimes {\cal V}_{\lambda},
\end{align}
where 
${\cal V}_{\lambda}$ is the representation space for 
the permutation group $S_n$,
${\cal U}_{\lambda}$ is the representation space for
the unitary group $U({H}_B)$.
\fi

\begin{proof}
The consequence \eqref{XH9} follows from \eqref{XH1} and the simple facts
\begin{align*}
D_\epsilon^h(
\{\mathcal{N}_1^{\otimes n}(R)\}_R\|
\{\mathcal{N}_2^{\otimes n}(S)\}_S)
&\le 
D^h_\epsilon(
\{\mathcal{N}_1^{\otimes n}(\rho^{\otimes n})\}_{\rho}\|\{\mathcal{N}_2^{\otimes n}(\sigma^{\otimes n})\}_{\sigma}), \\
\inf_{R,S}\frac{1}{n}D(
\mathcal{N}_1^{\otimes n}(R)\|
\mathcal{N}_2^{\otimes n}(S))
&\le 
\inf_{\rho,\sigma}
D(\mathcal{N}_1(\rho)\|\mathcal{N}_2(\sigma)) .
\end{align*}
By Theorem \ref{thm:adversarial}, one direction of \eqref{XH1} is easy
\begin{align}
\lim_{n\to \infty}\frac{1}{n}
D^h_\epsilon(
\{\mathcal{N}_1^{\otimes n}(R)\}_{R}\|\{\mathcal{N}_2^{\otimes n}(S)\}_{S})=D^{\inf, \infty }(\N_1\|\N_2)\le D^{\infty }(\N_1\|\N_2)
\label{XH1}
\end{align}
We now show the converse direction
\begin{align}
\inf_{\rho,\sigma}
D(\mathcal{N}_1(\rho)\|\mathcal{N}_2(\sigma)) 
\le
\lim_{n\to \infty}\frac{1}{n}
D^h_\epsilon(
\{\mathcal{N}_1^{\otimes n}(R)\}_{R}\|\{\mathcal{N}_2^{\otimes n}(S)\}_{S}) .\label{XH2}
\end{align}
Choose $r$ such that
\begin{align}
r<
\inf_{\rho,\sigma}
D(\mathcal{N}_1(\rho)\|\mathcal{N}_2(\sigma)) 
=\inf_{p,p'}
D\left(\sum_{x\in {\cal X}} p(x) \rho_{1,x}\| \sum_{x\in {\cal X}} p'(x) \rho_{2,x}\right)
\label{BH1}.
\end{align}
Using \cite[Eq. (26)]{MH}, 
we have
\begin{align}
\varepsilon(\{\mathcal{N}_1^{\otimes n}(R)\}_R\|
\{e^{nr}\mathcal{N}_2^{\otimes n}(S)\}_S)
=
\varepsilon(\{\mathcal{N}_1^{\otimes n}(R)\}_{R: \text{perm-inv}}\|
\{e^{nr}\mathcal{N}_2^{\otimes n}(S)\}_{S: \text{perm-inv}}).
\label{XH3}
\end{align}
where in the R.H.S. $R,S$ are over all  permutation-invariant states in $\mathcal{D}(H_A^{\ten n})$.
When $S$ is permutation-invariant, 
the distribution $\langle \vec{x}| S|\vec{x}\rangle$
is a permutation-invariant distribution.
Hence, $\mathcal{N}_2^{\otimes n}(S)$
is written as
\begin{align*}
\sum_{p \in T_d^n} q_2(p)
\sum_{\vec{x} \in {\cal X}^n}
P_p(\vec{x}) \rho_{2,\vec{x}}
=
\sum_{p \in T_d^n} q_2(p)\rho_{2,p}^{(n)},
\end{align*}
where
$\rho_{k,p}^{(n)}:=\sum_{\vec{x} \in {\cal X}^n}
P_p(\vec{x})\rho_{k,\vec{x}}$.
Similarly, $\mathcal{N}_1^{\otimes n}(R)$
is written as
\begin{align*}
\sum_{p \in T_d^n} q_1(p)
\sum_{\vec{x} \in {\cal X}^n}
P_p(\vec{x})
\rho_{1,\vec{x}}
=\sum_{p \in T_d^n} 
q_1(p)\rho_{1,p}^{(n)}.
\end{align*}

Thus, using \cite[Eq.(80)]{FH}, we have
\begin{align}
\varepsilon(\{\mathcal{N}_1^{\otimes n}(R)\}_{R: \text{perm-inv}}\|
\{e^{nr}\mathcal{N}_2^{\otimes n}(S)\}_{S: \text{perm-inv}})
=
\varepsilon(\{\rho_{1,p}^{(n)}\}_{p\in T_d^n}\|
\{e^{nr}\rho_{2,p'}^{(n)}\}_{p'\in T_d^n})
\label{XH4}
\end{align}

Also, since
\begin{align*}
\rho_{k,p}^{(n)}\le |T_d^n| 
\left(\sum_{x\in {\cal X}} p(x) \rho_{k,x}\right)^{\otimes n},
\end{align*}
 we have
\begin{align}
&\varepsilon(\{\rho_{1,p}^{(n)}\}_{p\in T_d^n}\|
\{e^{nr}\rho_{2,p'}^{(n)}\}_{p'\in T_d^n})
\\ \le &
\varepsilon\left(\left\{|T_d^n| \left(\sum\nolimits_{x} p(x) \rho_{1,x}\right)^{\otimes n} \right\}_{p\in T_d^n} \Big\|
\left\{e^{nr}|T_d^n| \left(\sum\nolimits_{x} p'(x) \rho_{2,x}\right)^{\otimes n}\right\}_{p'\in T_d^n}\right)
\label{XH5}.
\end{align}
Using \eqref{BAG1},we have
\begin{align}
&\varepsilon\left(\left\{|T_d^n| \left(\sum\nolimits_{x} p(x) \rho_{1,x}\right)^{\otimes n} \right\}_{p\in T_d^n} \Big\|
\left\{e^{nr}|T_d^n| \left(\sum\nolimits_{x} p'(x) \rho_{2,x}\right)^{\otimes n}\right\}_{p'\in T_d^n}\right)
\nonumber \\
&\le 
\sqrt{2}(2|T_d^n|)^2|T_d^n| 
\sum_{p, p'\in T_d^n}  
\left(\min_{s\in[0,1]}
e^{s r}
\tr \left[
\left(\sum\nolimits_{x} p(x) \rho_{1,x}\right)^{1-s}
\left(\sum\nolimits_{x} p'(x) \rho_{2,x}\right)^{s}  \right] \right)^{\frac{n}{2}} 
\nonumber\\
&\le 
4\sqrt{2}|T_d^n|^3
\sup_{p, p'\in T_d^n}  
\exp\left( -n \max_{s\in[0,1]} \frac{s}{2}\left(D_{1-s}(\sum\nolimits_{x} p(x) \rho_{1,x}
\big\| \sum\nolimits_{x} p'(x) \rho_{2,x})-r\right)\right) \nonumber\\
&\le 
4\sqrt{2}|T_d^n|^3
\exp\left( -n \max_s
\inf_{p, p'}  
 \frac{s}{2}\left(D_{1-s}\left(\sum\nolimits_{x} p(x) \rho_{1,x}
\big\| \sum\nolimits_{x} p'(x) \rho_{2,x}\right)-r\right) \right)
\label{XH6}.
\end{align}

By \eqref{BH1}, \eqref{XH6} and the cardinality of $|T_d^n|$ types is polynomial in $n$,
we have
\begin{align}
\varepsilon\left(\left\{|T_d^n| \left(\sum\nolimits_{x} p(x) \rho_{1,x}\right)^{\otimes n} \right\}_{p\in T_d^n} \Big\|
\left\{e^{nr}|T_d^n| \left(\sum\nolimits_{x} p'(x) \rho_{2,x}\right)^{\otimes n}\right\}_{p'\in T_d^n}\right)
\to 0,
\end{align}
which implies the following by combining 
\eqref{XH3}, \eqref{XH4} and \eqref{XH5},
\begin{align}
\varepsilon(\{\mathcal{N}_1^{\otimes n}(R)\}_R\|
\{e^{nr}\mathcal{N}_2^{\otimes n}(S)\}_S)
\to 0.\label{NM}
\end{align}
Therefore, for every $\epsilon>0$, we have
\begin{align}
\lim_{n}\frac{1}{n}D_\epsilon^h(\{\mathcal{N}_1^{\otimes n}(R)\}_R\|
\{\mathcal{N}_2^{\otimes n}(S)\}_S)
\ge r, 
\label{XH10}
\end{align}
which implies \eqref{XH2} as this holds for every $r$ below $\inf_{\rho,\sigma}
D(\mathcal{N}_1(\rho)\|\mathcal{N}_2(\sigma))$.
\end{proof}

\section{Adversarial Hypothesis Testing for CQ Channels}
\label{sec:CQ}
In this section, we analyze the adversarial hypothesis testing for CQ channels. Similar to the QQ case, we can consider two settings. We start with the case that Bob is informed with channel input. 
\subsection{Setting I: Bob informed}
In the adversarial setting, it is natural to assume that {\bf the receiver Bob can access the deterministic input signal $x\in \mathcal{X}$ which has been sent}. Hence, the effective channel on Bob's side is
\[\tilde{W}: p \mapsto \sum_{x\in\mathcal{X}} p(x)\,|x\rangle\!\langle x|\otimes \rho_{x} \ ,\]
where the outputs are CQ states on $\mathbb{C}^{|\mathcal{X}|}\ten \mathcal{H}$.  Then when the Alice send a state $W(x)$ with the distribution $x\sim p$ , Bob can choose a test $T_x$ conditional on the signal $x$ received. Effectively, this means a joint test $$T=\sum_{x} |x\rangle\!\langle x|\otimes T_{x}\in \mathcal{B}(\mathbb{C}^{|\mathcal{X}|}\ten \mathcal{H}) \ ,\  0\le T_x\le I \text{ for all } x.$$ 

For two CQ channels $W_1$ and $W_2$, the optimal error probability is equivalent to consider the effective channel $\tilde{W}_1$ and $\tilde{W}_2$ with input distribution $p$ not informed. In this case, Bob discriminate between two state sets $\{ \tilde{W}_1(p)\}_p$ and $\{ \tilde{W}_2(q)\}_q$
\[ \beta_\varepsilon(\{\tilde{W}_1(p)\}_{p}, \{\tilde{W}_2(q)\}_{q})= \inf_{0\leq T \leq I} \left\{ \max_{q}\tr (T\tilde{W}_2(q)) : \max_{p}\tr ((I-T)\tilde{W}_1(p)))\le \varepsilon  \right\}, \]
where the maximum are over $p,q\in \mathcal{P}(\mathcal{X})$. It turns out that for effective channels, this is equivalent to consider the input distribution $p$ is informed.
\begin{align*} \beta_\varepsilon(\tilde{W}_1, \tilde{W}_2)=& \max_{p} \inf_{0\leq T\leq I} \{ \tr (T\tilde{W}_2(p)) : \tr ((I-T)\tilde{W}_1(p)))\le \varepsilon \} \end{align*}
\begin{lemma}
Let $W_1,W_2$ be two CQ channels and $\tilde{W}_1,\tilde{W}_2$ be the effective channels defined above. Then for any $\varepsilon\in (0,1)$,
\[ \inf_{p\in \mathcal{P}(\mathcal{X})}D_h^\varepsilon(\tilde{W}_1(p)\|\tilde{W}_2(p))=D_h^\varepsilon(\{\tilde{W}_1(p)\}_{p}\|\{\tilde{W}_2(q)\}_{q})=\inf_{x\in \mathcal{X}}D_h^\varepsilon(\rho_{1,x}\|\rho_{2,x}) 
\]
\end{lemma}
\begin{proof}
Given the CQ structure of the output states $\tilde{W}_1(p)$ and tests T, we have
\begin{align*}
 \beta_\varepsilon(\tilde{W}_1, \tilde{W}_2)= &\max_{p} \inf_{0\leq T_x\leq I}\left\{ \sum_{x} p(x)\tr(T_x\rho_{2,x}) :\sum_{x} p(x)\tr(T_x\rho_{1,x})\ge 1- \varepsilon \right\} 
 \\ \le &  \max_{p} \inf_{0\leq T_x\leq I} \left\{ \sum_{x} p(x)\tr(T_x\rho_{2,x}) :\tr(T_x\rho_{1,x})\ge 1- \varepsilon \ \forall\  x\in\mathcal{X} \right\} 
 \\  = & \max_{x} \inf_{0\leq T_x\leq I} \left\{ \tr(T_x\rho_{2,x}):\tr(T_x\rho_{1,x})\ge 1- \varepsilon \right\} 
 \\ = & \max_{x}  \beta_\varepsilon(\rho_{1,x}, \rho_{2,x})\le  \beta_\varepsilon(\tilde{W}_1, \tilde{W}_2)
\end{align*}
where the last inequality follows from choosing point mass. Hence, 
\[ \beta_\varepsilon(\tilde{W}_1, \tilde{W}_2)=\max_{x\in\mathcal{X}}\beta_\varepsilon(\rho_{1,x}, \rho_{2,x})\ .\]

For the case $p$ is not informed,  by the CQ structure of the output states and tests again, we have
\begin{align}
 \beta_\varepsilon(\{\tilde{W}_1(p)\}_{p}, \{\tilde{W}_2(q)\}_{q})= &\inf_{0\le T_x\le I} \left\{ \max_{q}\sum_{x} q(x)\tr(T_x\rho_{2,x}) : \min_{p}\sum_{x'}p(x')\tr(\rho_{1,x'}T_{x'})\ge  1-\varepsilon  \right\} 
\nonumber \\ = &  \inf_{0\le T_x\le I}\left\{ \max_{x}\tr(T_x\rho_{2,x}) :\tr(T_{x}\rho_{2,x})\ge 1- \varepsilon \ \forall \ x\in \mathcal{X}\right\} 
\nonumber \\  = & \max_{x} \left\{ \tr(T_x\rho_{2,x}):\tr(T_x\rho_{1,x})\ge 1- \varepsilon \ \forall \ x\in \mathcal{X}\right\} 
 \nonumber\\ = & \max_{x}  \beta_\varepsilon(\rho_{1,x}, \rho_{2,x})=\beta_\varepsilon(\tilde{W}_1, \tilde{W}_2)\label{eq:eq}
\end{align}
which has the same error as the previous case.
\end{proof}

The above equality \eqref{eq:eq} has the meaning that if Bob can access the deterministic input signal $x\in \mathcal{X}$ being applied, then it does not matter whether Alice's input distribution $p$ is informed to him, because choosing the test operator $T_x$ depending on the input actual signal $x$ is more powerful than choosing the test operator $T_p$ depending on the input probability distribution $p$. In this case, the optimal input for Alice (to maximize the error) is a deterministic signal $x$, which in the setting will be accessible for Bob.

In the $n$-shot setting, if Alice's input to the $n$-fold effective channel is arbitrary general probability distribution $P$ on $\mathcal{X}^n$ (let's denote the set of distribution as $\mathcal{P}(\mathcal{X}^n)$), the output at Bob is 
\begin{align} 
    \tilde{W}_i^{\otimes n}(p^n)
    &= \!\sum_{\vec{x}\in\mathcal{X}^n} \! p^n\!(\vec{x}) |\vec{x}\rangle\langle \vec{x}|\otimes \rho_{i,x_1} \otimes \cdots \otimes \rho_{i,x_n},  \label{eq:tilde_n-shot}
\end{align}
where $\vec{x}= x_1 x_2 \cdots x_n$ is a $n$-sequence in $\mathcal{X}^n$.
By the above lemma, we are interested in the error probability
\begin{align} \label{eq:CQ_informed_general}
D_h^{\epsilon}\big( \{\tilde{W}_1^{\otimes n}(P)\}_{P\in \mathcal{P}(\mathcal{X}^n)} \| \{\tilde{W}_2^{\otimes n}(Q) \}_{Q\in\mathcal{P}(\mathcal{X}^n)}\big)=\frac{1}{n}
\inf_{\vec{x}\in \mathcal{X}^n} D_h^\varepsilon( \rho_{1,\vec{x}}^{(n)}\| \rho_{2,\vec{x}}^{(n)})
\end{align}
Note that we are discriminating two sets of output states because Bob does not know the distribution $p^n \in \mathcal{P}(\mathcal{X}^n)$ that Alice is using.

We show that the Stein exponent in this setting is single-letter, which coincides with the case when Alice uses only i.i.d.~input $p^{\otimes n}$ or even identical deterministic inputs $ (x, \ldots, x)$, $x\in\mathcal{X}$. 
\begin{theorem}[Adversarial hypothesis testing for CQ channels]\label{thm:cqinform}
Let $W_1,W_2$ be two CQ channels and $\tilde{W}_1,\tilde{W}_2$ be the effective channels defined above. We have
\begin{align}
\lim_{n\to\infty}\frac{1}{n}
 D_\varepsilon^h( \{\tilde{W}_1^{\ten n}(p^n)\}_{p^n}\| \{\tilde{W}_2^{\ten n}(q^n))\}_{q^n})=\lim_{n\to\infty}\frac{1}{n}
 D_\varepsilon^h( \{\tilde{W}_1^{\ten n}(p^{\ten n})\}_{p}\| \{\tilde{W}_2^{\ten n}(q^{\ten n})\}_{q})=\inf_{x\in \mathcal{X}} D(\rho_{1,x}\|\rho_{2,x}).
\end{align}
\end{theorem}
\begin{proof}
Let ${\N}_1,{\N}_2$ be the EB channels corresponding to the effective channel $\tilde{W}_1,\tilde{W}_2$ as follow
\begin{align*}
{\N}_{i}(\sigma)=\sum_{x}\bra{x}\sigma\ket{x} \ketbra{x}\ten \rho_{i,x}\  ,\ i=1,2
\end{align*}
Note that for any $n$, ${\N}_i^{\ten n}$ and $\tilde{W}_i^{\ten n}$ has the same :  
\begin{align*} &\{{\N}^{\ten n}_{i}(R)\}_{R\in \mathcal{D}(\H^{\ten n})}=\{\tilde{W}_i^{\otimes n}(p^n)\}_{p^n\in \mathcal{P}(\mathcal{X}^n)},  \\
&\{{\N}^{\ten n}_{i}(\rho^{\otimes n})\}_{\rho\in \mathcal{D}(\H)}=\{\tilde{W}_i^{\otimes n}(p^{\ten n})\}_{p\in \mathcal{P}(\mathcal{X})}.  
\end{align*}
Using Stein's exponent for the EB channels in Theorem \ref{thm:EB}, we have
\begin{align*}
&\lim_{n\to\infty}\frac{1}{n}
 D_\varepsilon^h( \{\tilde{W}_1^{\ten n}(p^n)\}_{p^n}\| \{\tilde{W}_2^{\ten n}(q^n))\}_{q^n})
 =\lim_{n\to\infty}\frac{1}{n}
 D_\varepsilon^h( \{{\N}^{\ten n}_{1}(R)\}_{R}\|  \{{\N}^{\ten n}_{2}(S)\}_{S})\\
  =&\lim_{n\to\infty}\frac{1}{n}
 D_\varepsilon^h( \{\tilde{W}_1^{\ten n}(p^{\ten n})\}_{p}\| \{\tilde{W}_2^{\ten n}(q^{\ten n})\}_{q})=\lim_{n\to\infty}\frac{1}{n}
 D_\varepsilon^h( \{{\N}^{\ten n}_{1}(\rho^{\otimes n})\}_{\rho}\| \{{\N}^{\ten n}_{2}(\sigma^{\otimes n})\}_{\sigma})\\
 = &\inf_{\rho,\sigma}D({\N}_1(\rho)\|{\N}_2(\sigma))\\
 =&\inf_{p,q}D(\tilde{W}_1(p)\|\tilde{W}_2(q))
\end{align*}
We calculate that
\begin{align*}
&D(\tilde{W}_1(p) \| 
\tilde{W}_2(q))
= D\left(\sum_{x}p(x)\ketbra{x}\ten \rho_{1,x} \| 
\sum_{x}q(x)\ketbra{x}\ten \rho_{2,x}\right)=
D(p\|q)+ \sum_x p(x) D(\rho_{1,x}\|\rho_{2,x}) .
\end{align*}
Taking infimum over $p,q\in \mathcal{P}(X)$, we have
$ \displaystyle\inf_{p,q}D(\tilde{W}_1(p)||\tilde{W}_2(q))=\inf_{x}D(\rho_{1,x}\|\rho_{2,x}).$
\end{proof}
\begin{rem}{\rm The above exponent can also be obtained by the informed case exponent $D^\infty(\tilde{W}_1\|\tilde{W}_2)$ in Theorem \ref{thm:adversarial}.
Indeed, note that for effective channels $\tilde{W}_1,\tilde{W}_2$
\begin{align}D(\tilde{W}_1\|\tilde{W}_2)=&\inf_{p}D(\sum_{x}p(x)\ketbra{x}\ten \rho_{1,x} \| 
\sum_{x}p(x)\ketbra{x}\ten \rho_{2,x})\\ =&\inf_{p}\sum_x p(x) D(\rho_{1,x}\|\rho_{2,x}) =\inf_{x}D(\rho_{1,x}\|\rho_{2,x})
\end{align}
In particular, $D(\tilde{W}_1\|\tilde{W}_2)=D^{\inf}(\tilde{W}_1\|\tilde{W}_2)$.
It is easy to verify the additivity
\[D(\tilde{W}_1\ten \tilde{V}_1\|\tilde{W}_2\ten \tilde{V}_2)=D(\tilde{W}_1\|\tilde{W}_2)+D(\tilde{V}_1\|\tilde{V}_2) \]
Then we have 
\begin{align}
D^\infty(\tilde{W}_1\|\tilde{W}_2)=D(\tilde{W}_1\|\tilde{W}_2)=\inf_{x}D(\rho_{1,x}\|\rho_{2,x})
\end{align}
which also gives the adversarial Stein's exponent. Moreover, this shows that $\inf_{x}D(\rho_{1,x}\|\rho_{2,x})$ is also the exponent for Alice using: general sequence $\vec{x}=x_1\cdots x_n$, or  identical sequence $x^n=x\cdots x$,
\begin{align}
\inf_{x\in \mathcal{X}} D(\rho_{1,x}\|\rho_{2,x})=\lim_{n\to\infty}\frac{1}{n}\inf_{\vec{x}}
 D^\varepsilon_h( \rho_{1,\vec{x}}\| \rho_{2,\vec{x}})=\lim_{n\to\infty}\frac{1}{n}
 D^\varepsilon_h( \rho_{1,x}^{\ten n}\| \rho_{2,x}^{\ten n})
\end{align}
where $\rho_{i,\vec{x}}=\rho_{i,x_1} \otimes \cdots \otimes \rho_{i,x_n}$ is the product output state of sequence $\vec{x}=x_1\cdots x_n$. 
}
\end{rem}

\subsection{Setting II: Bob non-informed}
For non-informed case  of CQ channel, we consider the case Bob cannot see 
any thing except for the output state $W_i(p)$. More precisely, we assume {\bf Bob is neither informed the deterministic signal $x$ being sent or the input distribution $p$ Alice uses}.
Hence, this setting is the equivalent to 
the non-informed case of the corresponding EB channel, which is different from the above effective channel. By simply applying Theorem \ref{thm:EB} for the non-informed case of the corresponding EB channel, we have that for both general input and i.i.d.~input.

\begin{theorem} \label{thm:cqnoninform}
Let $W_1,W_2$ be two CQ channels.
Then for any $\varepsilon\in (0,1)$,    
\begin{align}
    &\lim_{n\to\infty} \frac{1}{n} D_h^{\epsilon}\big( \{{W}_1^{\otimes n}(p^{n})\}_{p^n} \| \{{W}_2^{\otimes n}(q^{n}) \}_{q^n}\big)
    \notag
    \\
    &= \lim_{n\to\infty} \frac{1}{n} D_h^{\epsilon}\big( \{{W}_1^{\otimes n}(p^{\otimes n})\}_p \| \{{W}_2^{\otimes n}(q^{\otimes n})\}_q \big)
    \\
    &= \inf_{p,p'}D\left(\sum\nolimits_{x}p(x)\rho_{1,x}\|\sum\nolimits_{x}p'(x) \rho_{2,x}\right),
\end{align}
where the infimum is over $p,q\in \mathcal{P}(\mathcal{X})$; and $p^n, q^n \in \mathcal{P}(\mathcal{X}^n)$.
\end{theorem}

\section{Discussions} \label{sec:discussions}
Let $W_1,W_2$ be two CQ channels and $\N_1,\N_2$ be the corresponding EB channels 
\[\N_i(\sigma)=\sum_{x}\bra{x}\sigma\ket{x}\rho_{i,x}\ ,\ i=1,2.\]
By Theorem \ref{thm:cqinform} 
in the informed setting, the adversarial Stein's exponent for the CQ channels $W_1,W_2$ are
\begin{align}
&\lim_{n\to\infty}\frac{1}{n}\log D_h^\varepsilon( \tilde{W}_1^{\ten n}, \tilde{W}_2^{\ten n})\nonumber\\
= &\lim_{n\to\infty}\frac{1}{n}\log D_h^\varepsilon( \{\tilde{W}_1^{\ten n}(P)\}_P, \{\tilde{W}_2^{\ten n}(Q)\}_Q)\nonumber\\
=  &\lim_{n\to\infty}\frac{1}{n}\log D_h^\varepsilon( \{\tilde{W}_1^{\ten n}(p^{\ten n})\}_n, \{\tilde{W}_2^{\ten n}(q^{\ten n})\}_q)   \nonumber\\
= &\inf_{x\in \mathcal{X}} D(\rho_{1,x}\|\rho_{2,x}) \tag{general \& i.i.d.}. 
\end{align}
By Theorem \ref{thm:adversarial}, the adversarial Stein's exponent for the corresponding EB channels $\N_1,\N_2$ are
\begin{align}
&\lim_{n\to\infty}\frac{1}{n}\log \inf_{R}D_h^\varepsilon( \N_1^{\ten n}(R)\|\N_1^{\ten n}(R))=D^{\inf}(\N_1\|\N_2) \nonumber
\\
= &\inf_{p,p'\in \mathcal{P}(\mathcal{X})} D\left(\sum_{x}p(x)\rho_{1,x}\|\sum_{x}p'(x)\rho_{2,x}\right)=D^{\inf}(W_1\|W_2), \tag{general}
\\
&\lim_{n\to\infty}\frac{1}{n}\log \inf_{\rho}D_h^\varepsilon( \N_1^{\ten n}(\rho^{\ten n})\| \N_2^{\ten n}(\rho^{\ten n}))=D(\N_1\|\N_2)\nonumber\\
= &\inf_{p\in \mathcal{P}(\mathcal{X})} D\left(\sum_{x}p(x)\rho_{1,x}\|\sum_{x}p(x)\rho_{2,x}\right)
=D(W_1\|W_2). \tag{i.i.d.}
\end{align}
In general, these exponent can be different 
\begin{align} 
&\inf_{x\in \mathcal{X}} D(\rho_{1,x}\|\rho_{2,x}) >D^{\inf}(W_1\|W_2), \tag{general}
\\
&\inf_{x\in \mathcal{X}} D(\rho_{1,x}\|\rho_{2,x}) >D(W_1\|W_2), \tag{i.i.d.}
\end{align}
as in the Example \ref{exam:1}. 
See also Table~\ref{table:CQ_vs_EB}.

This also shows that the adversarial hypothesis testing of CQ channel is different with corresponding EB channels. Indeed, given the assumption that the receiver Bob can access the actual input signal signal $x$ being set, the adversarial hypothesis testing for CQ channels has better performance than for the corresponding EB channels. For EB channel, such assumption cannot be made because Alice can use super-position pure state $\sum_{x}\sqrt{p(x)}\ket{x}$ to produce the randomized the output states $\sum_{x}p(x)\rho_x$, where the randomness from the measurement effect can not be obtained by Bob.

\bibliography{adversarial, reference}
\bibliographystyle{myIEEEtran}
\end{document}